# Prediction Analysis of Optical Tracker Parameters using Machine Learning Approaches for efficient Head Tracking

AMAN KATARIA[1], SMARAJIT GHOSH[1], VINOD KARAR[2]

[1] Department of Electrical and Instrumentation Engineering, Thapar University, Patiala-147004, Punjab, India
[2] Department of Optical Devices and Systems, CSIR-Central Scientific Instruments Organization, Chandigarh, India

A head tracker is a crucial part of the head mounted display systems, as it tracks the head of the pilot in the plane/cockpit simulator. The operational flaws of head trackers are also dependent on different environmental conditions like different lighting conditions and stray light interference. In this letter, an optical tracker has been employed to gather the 6-DoF data of head movements under different environmental conditions. Also, the effect of different environmental conditions and variation in distance between the receiver and optical transmitter on the 6-DoF data was analyzed.

**Keywords**: Machine Learning, Head Tracking, Random Forest, Optical Head Tracking, Aviation

**Introduction:** Head tracking has always been a significant part in designing of Head mounted display systems (HMDs). HMDs are considered as integral part of fighter aircraft avionics. A head tracker is a crucial part of the HMDs, as it tracks the head of the pilot in the plane/cockpit simulator to synchronize the view of outer world with the view in the HMDs. There are various types of head trackers which work with different tracking techniques. Their operational flaws are also dependent on different environmental conditions which are different lighting conditions and stray light interference.

Tracking can be positional or orientation in terms of coordinates and when both are combined, the system is recognized as six degrees of freedom (6-DoF) system [1]. Directions of head tracking in a 6 DOF system are X, Y, and Z known as positional coordinates and Roll, Yaw and Pitch known as orientation coordinates. TrackIR™ 5 Optical head tracker is used to track the six DOF directions of head movement. TrackIR™ 5 optical tracker acts as a source to generate the Infrared (IR) light. The incoming IR light is reflected back by the three retro-reflective markers, which are assembled in the TrackClip which is placed on the helmet/head of the pilot. The reflected IR light is detected by the infrared camera present in TrackIR™ 5, which then tracks the positional and orientation coordinates of the head of the pilot.

Experiments were conducted under different light intensities (LI) and different distance range values between transmitter and receiver (tracker) on which the performance of an optical tracker strongly depends. Generally, the expected performance of an optical tracker diverges largely if the stray light interference is more. Each optical tracker has limited work capacity under some particular LI [2]. In the experiment, the 6-DoF data is predicted through different machine learning methods under different light intensity and distance variation between sensor and transmitter to virtually judge the performance of an optical tracker [3]. A desirable predictive method is required which can predict the 6-DoF data under different light intensity of the environment and distance between transmitter and receiver [4]. According to the variations in environmental conditions, appropriate number of trackers could be incorporated in a HMDs. Also in the aircraft simulator, it can help to choose the appropriate optical tracker according to the environmental condition of simulator bed.

Machine learning has been used to great extent in predicting mathematical and statistical optimization [5]. In the proposed work, four methods of machine learning applied to predict the 6-DoF data under different light intensities and the distance between transmitter (infrared light source) and receiver (optical receiver). Six DOF directions of head tracking are analyzed for different environmental conditions. For predicting the output, Random forest, Neural networks, Generalized linear model (GLM) and Support vector machine (SVM) are used. To determine the stableness and robustness of the best method of prediction, the K-fold cross validation is performed [6].

Remaining paper is structured with Section 1 presenting the experimental setup, Section 2 represents the methodology used. Model evaluation is described in section 3. Results are discussed in section 4. Finally conclusion and future scope is presented in section 5 and section 6 respectively.

1. **Experimental Setup:** For carrying out desired experimentation conditions, the Cockpit Simulator is used. It comprised of pilot seat at designated distance from the transmitter. The aircraft canopy covered the head up display along with the optical transmitter. The receiver (infrared retro-reflector) was mounted on the user cap. The diffused light source controlled the ambient light variations, while the pilot movement simulated the distances between the transmitter and the receiver. TrackIR 5™ is used to track the head movements in Display Simulator. The light intensity was measured through Lux Meter (Model TES 1332 Digital Lux Meter).

a) Display Simulator

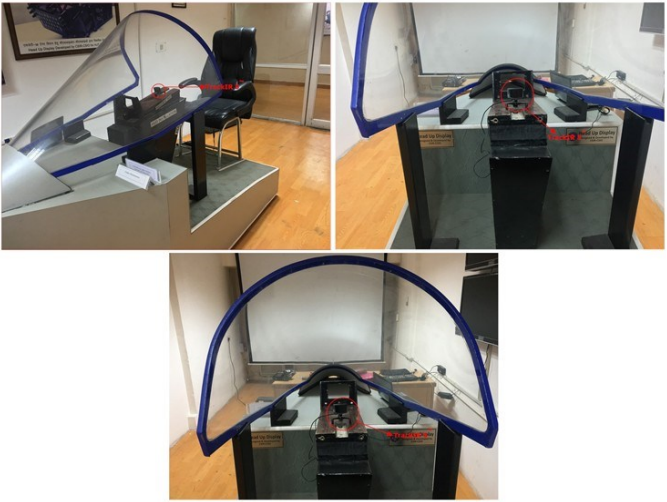

Fig.1. Display Simulator with optical tracker TrackIR 5™ (Shown in red circle) at CSIR- CSIO

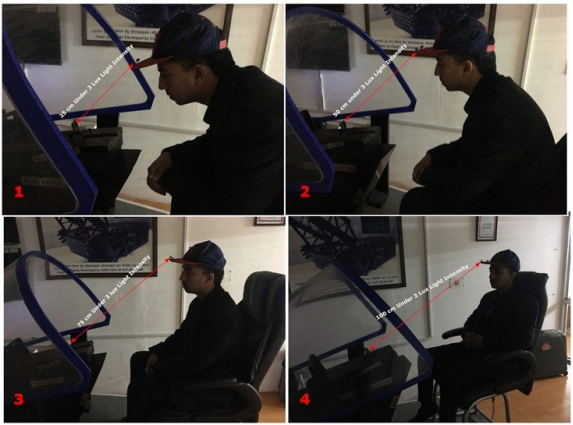

Fig.2. Experiment conducted under 3 lux light intensity with different distances 25cm, 50cm, 75cm and 100cm marked as 1, 2, 3 and 4 respectively.

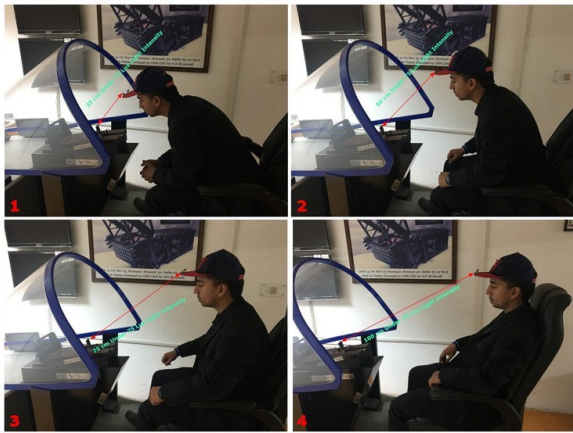

Fig.3. Experiment conducted under 75 lux light intensity with different distances 25cm, 50cm, 75cm and 100cm marked as 1, 2, 3 and 4 respectively

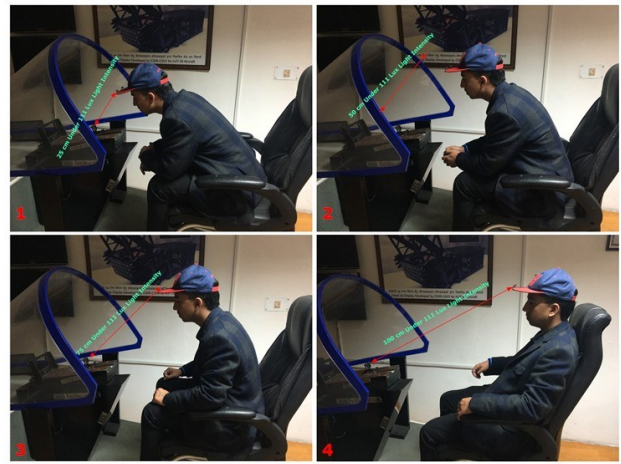

Fig.4. Experiment conducted under 111 lux light intensity with different distances 25cm, 50cm, 75cm and 100cm marked as 1, 2, 3 and 4 respectively.

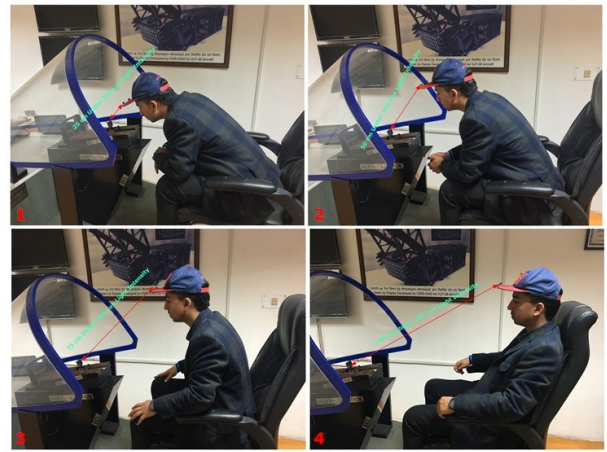

Fig.5. Experiment conducted under 165 lux light intensity with different distances 25cm, 50cm, 75cm and 100cm marked as 1, 2, 3 and 4 respectively.

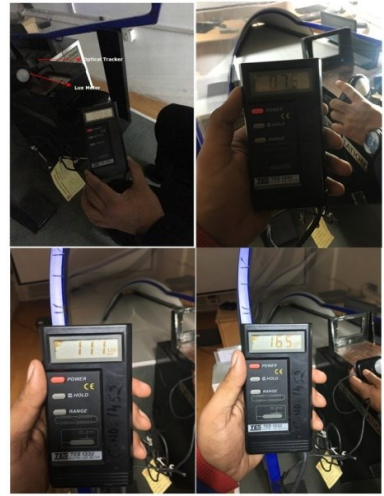

Fig.6 Lux meter readings under different light intensity.

## 2. Methodology Used

### 2.1 Description of Data Set and its Features

Data set have been modeled after performing experiments under different conditions viz. variation in light intensity and variation in distances between the transmitter and the receiver. Description regarding movements of head in optical head tracking used in the experiment has been described in Table I. Values of light intensity, distances between transmitter and receiver and sample values of 6-DOF head tracking coordinates are provided in Table- II.

Table- I: Description of the movements of head recorded by an Optical Tracker

| DoF | Information |
|---|---|
| X | Forward and backward translational coordinate of the pilot's head |
| Y | Left and right translational coordinate of the pilot's head |
| Z | Up and Down translational coordinate of the pilot's head |
| Yaw | Rotational coordinate of the pilot's head along Z-axis |
| Pitch | Rotational coordinate of the pilot's head along Y-axis |
| Roll | Rotational coordinate of the pilot's head along X-axis |

| Light intensity (in lux) | Distance (in cm) | Yaw | Pitch | Roll | X | Y | Z |
|---|---|---|---|---|---|---|---|
| 3 | 25 | -18.2 | 3.8 | -17.8 | -4.5 | 4.1 | 13.47 |
| 3 | 50 | -20.5 | -1.7 | -7.8 | 0.2 | 0.7 | -10.3 |
| 3 | 75 | -23.5 | -12.78 | -2.67 | -13.6 | -0.2 | 3.97 |
| 3 | 100 | 5.1 | -14.44 | -2.9 | 6 | 0.7 | -5.13 |
| 75 | 25 | -54.3 | 1.1 | -0.6 | -2.8 | 0.51 | 6.05 |
| 75 | 50 | -49 | -1.4 | -1.8 | -1.2 | 1.4 | 2.95 |
| 75 | 75 | -8.8 | 2 | -4.5 | -1.9 | 1.3 | -0.16 |
| 75 | 100 | -11.6 | 1 | -3.8 | 0.3 | -1.9 | -12.36 |
| 111 | 25 | -13.1 | -2.8 | 0.1 | -7.6 | 3.2 | 5.66 |
| 111 | 50 | -7.4 | -3 | 0 | -6.2 | 3.3 | 5.38 |
| 111 | 75 | -2 | -2.8 | -0.4 | -1.1 | 3 | 6.46 |
| 111 | 25 | -16 | -2.8 | -0.4 | 2 | 2.9 | -2.15 |
| 165 | 50 | 9.3 | -6.6 | 2.4 | 0.6 | 2.5 | -9.43 |
| 165 | 25 | 9.7 | -6.1 | 2 | 1 | 2 | -11.21 |
| 165 | 75 | 9.43 | -6.67 | 2.4 | 0.6 | 2.5 | -9.43 |
| 165 | 100 | -10.22 | -7.99 | -1.8 | 5 | 2 | -7.76 |

Table 2. Sample set of Data

### 2.2 Quantitative Assessment

Experiments conducted with light intensity and distance variation between the transmitter and the receiver helps in predication of the desirable conditions in which optical tracker give satisfactory output. Light intensity is measured as the rate at which energy of light is delivered per unit surface per unit time per unit area. Light intensity is measured in Lux (lx). The experiments are conducted for light intensity varying form 4 lux to 166 lux. In the experimentation, the distance between the pilot's head (transmitter is located on cap worn on head) and the receiver are kept in four distances: 25cm, 50cm, 75cm and 100cm.

### 2.3 Feature Description

Six physical DOF directions are extensively used in the experimentation and subsequent analysis. The X coordinate in this case is the translation of the body along X-axis also known as forward and backward translation. Forward translation leads to positive value and backward translation leads to negative values. The Y coordinate is the translation of the body along Y-axis also known as left and right translation. Right translation leads to positive value and left translation leads to negative values. Z coordinate is the coordinate of translation of the body along Z-axis also known as up and down translation. Up translation leads to positive value and down translation leads to negative value. The rotational coordinates are Yaw, Pitch and Roll. Yaw is the rotation along Z-axis and is measured in degrees. Roll is the rotation along X-axis and is measured in degrees and Pitch is the rotation along Y-axis and is measured in degrees. The rotational matrixes of the rotational coordinates are given below:

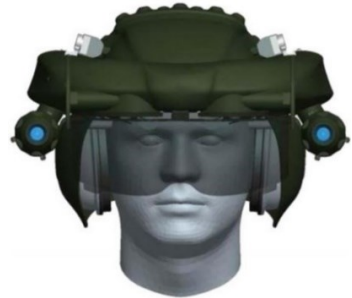

Fig.7 Prototype view of Optical Head tracker [7]

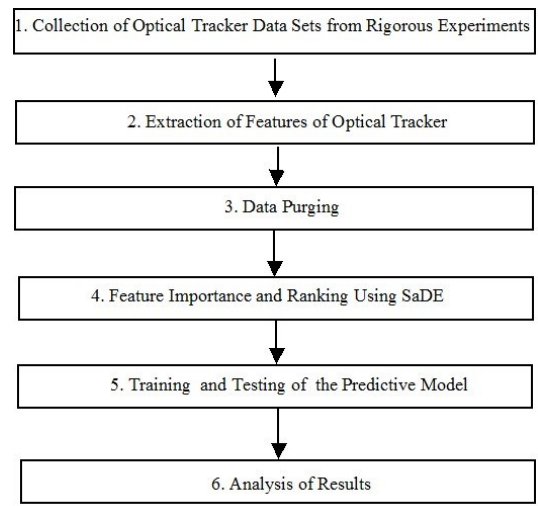

Fig.8 Methodology used

## 2.4 Approach

The experimental approach is presented in Figure 4. Data sets are collected by conducting experiments under different conditions: combinations of four distance ranges – 25 cm, 50 cm, 75 cm & 100 cm, and four light intensities - 4 lux, 75 lux, 110 lux and 166 lux. The lux values are measured at the receiver side using Lux Meter Model TES 1332 Digital Lux Meter. Next, elimination of duplicate and missing entries from data set is processed as elimination of the repetitive value assures the uniqueness in experimental data set using IBM SPSS/Microsoft Excel. To describe the importance of each feature in terms of ranking, Self adaptive differential evolution technique is used which is discussed in the next section. Next, five machine learning techniques, as shown in Table 4, are trained and then tested on the corresponding optical tracker data set with predefined parameters of machine learning techniques discussed in further sections.

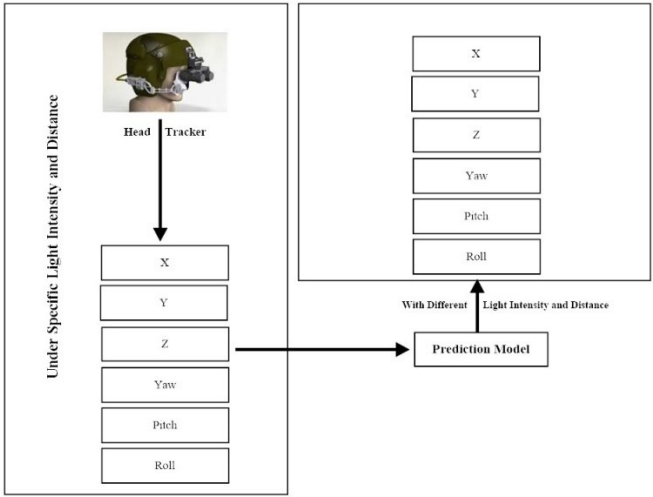

Fig.9 Prediction Model

Prediction model is described in Figure 9. In the final stage the evaluation of the method is done on Accuracy (eq. 4). Robustness of the best predictive machine learning method is determined using K-fold cross validation.

### 2.4.1 Self Adaptive Differential Evolution (SaDE)

SaDE as proposed by Qin et.al [8], is one of the efficient, fast and robust algorithm to solve many types of real optimization problems. In this technique, the controlling parameter G and CO, and the choice of learning strategy is not required to be predefined. Learning strategies along with parameters, which are associated with SaDE, can be adapted automatically during the procedure of evolving. Accordingly with SaDE, it is possible to find out more suitable generation scheme along with the parameter settings adaptively to cope up with different phases of the search evolution. Four efficient trial vectors generation schemes are used in SaDE: (i) DE/rand/1/bin, (ii) DE/rand-to-best/2/bin, (iii) DE/rand/2/bin, and (iv) DE/current-to-rand/1. These schemes are assigned to generate a scheme candidate pool. Binomial type crossover is used in (i), (ii) and (iii) differential evolution variants while arithmetic recombination is used in (iv) differential evolution variant. For every target vector existing in the current population, single trial vector scheme is selected from the candidate pool in accordance with the probability learned from its rate of success in generating improved solutions, which is also known as Learning period (LP). The selected scheme is then applied to the corresponding target vector to evolve the trial vector. In more generalized manner, the probabilities of selecting each scheme at every generation in the candidate pool are summed up to 1. At initial stage, these probabilities are equal (1/P for P total number of schemes in pool), which are then adapted gradually during the evolution, which is further based upon the success and failure rates of previous LP evolutions. The probabilities adaptations occurs in such pattern that the larger is the rate of success for $p^{th}$ scheme in the pool among previous LP generations, the larger is the probability which is applied to generate trial vectors at present generation [9].

### 2.4.2 Importance of Feature using SaDE

According to the objective function defined in equation 1 for light intensity and Distance (D), SaDE assigns an optimum weight to each feature of optical tracker. The mutation rate (MR) and crossover rate (CR) used in this algorithm are set to be 0.01 and 0.9 respectively.

$$\text{Objective function} = min\left(\sum_{m=1}^{R}\sqrt{\left(T_i - \sum_{s=1}^{p} w_s . P_{m,s}\right)^2}\right) \quad (1)$$

where, R is the total number of instances in training data set, T will be LI and D target, P is physical parameters (X, Y, Z, Pitch, Roll and Yaw), p is the number of features or properties (6 in this case) and w is the weight assigned to every parameter defined in the domain [0,1]. After five different instances, the description of the average weight assigned to each feature as also shown in Table 3 for LI and D range. The average weight obtained is used to rank the features. According to SaDE algorithm, it is found that Pitch has the highest ranking and Roll has the lowest ranking. Ranking based on perturbations caused in 6-DoF data due to different light intensities and distance between transmitter and receiver may help in selecting the appropriate optical tracker or other tracking technology tracker like magnetic or inertial head tracker.

|  | 6 –DoF Movements | | | | | |
|---|---|---|---|---|---|---|
|  | Pitch | Y | X | Z | Yaw | Roll |
| Different Distance | 79.7 | 74.4 | 60.1 | 75.1 | 5.9 | 49.6 |
| Different Light Intensity | 83.3 | 80.0 | 62.5 | 82.2 | 70.2 | 47.6 |
| Average | 81.5 | 77.2 | 61.3 | 78.7 | 68.1 | 48.6 |
| Perturba-tion Ranking | 1 | 2 | 5 | 3 | 4 | 6 |

Table 3. Perturbation ranking of the 6-DoF data of TrackIR Optical tracker using SADE

## 2.5 Methods of Machine learning

In this experiment, four different machine learning techniques as shown in Table 4, are used for predicting 6-DoF data under different LI and Distance range *betwen* pilot's head (transmitter) and optical tracker (receiver). For prediction, machine learning methods used in this experiment are available on open source software known as R licensed under GNU GPL. Methods of machine learning used are briefly discussed below:

1. Random Forest (randomForest): Random forest is a learning method used for classification based on the technique forest of trees utilizing the random inputs [10]. They can map non linear relationships unlike linear model. It is also known as non parametric method, which needs no assumptions about the distribution of data.

2. Linear Models (GLM): Linear models allow for response variables, which have models for error distribution other than the normal distribution. For analysis of covariance, this method uses linear models. This method also carries out single stratum analysis of variance and regression [12]. This model lays emphasis on linear data, so if the data is linear in nature it will yield high accuracy.

3. Support Vector Machine (SVM): SVMs are supervised learning models with learning algorithms, which analyze data for classification analysis. SVMs are effective and robust techniques for nonlinear classification and regression [13]. SVM is used to compare the results with the other machine learning methods; however, it is less efficient method for the training purposes.

4. Neural Network (NN): Neural Networks are Von Neumann machines, which are based on abstraction of information processing. Using back propagation algorithm with or without weight, training of neural networks is done [14]. NN are used because of their ability to detect the complex non linear relationships between the dependent and independent variables within the data set, but it yield less accuracy in this experiment due to slow computational time and proneness to over-fitting.

| Model | Package | Parameter setting | Reference |
|---|---|---|---|
| R Forest | mtry | Number of tree = 500 | [14] |
| LM | stats | Multimonial | [11] |
| SVM | e1071 | Epsilon = 0.5; nu = 10 | [12] |
| NN | neuralnet | MaxNWTs = 10000; hlayers = 10; maxit = 100 | [13] |

Table 4 describes the machine learning packages used in open source software R with the predefined settings and package.

## 3 Model Evaluation

Performance of the prediction can be measured in various ways, where one way can be more desirable than other ways which depends on the application evaluated. In this paper, by applying same number of input variables (i.e. features of optical head tracking) two different models for the prediction of two output variables (i.e. light intensity and distance between transmitter and tracker) has been developed. The function formulated, which can be used in all the machine learning models, is described by:

Light Intensity ~ f (X, Y, Z, Roll, Yaw, Pitch)     (2)

Distance ~ f (X, Y, Z, Roll, Yaw, Pitch)     (3)

The accuracy in this experiment is the percentage in the deviation of predicted target from the actual target.

$$\text{Accuracy} = \frac{100}{z}\sum_{s=1}^{z} f_i \qquad f_i = 1 \text{ if } \{t_i = x_i\}$$

$$0 \text{ otherwise} \quad (4)$$

Where x is real target, t is predicted target, and z is the total number of instances.

## 4 Results

Prediction results obtained from all the four machine learning methods have been analyzed in this Section. Training, testing and validation have been performed through different machine learning methods. In training-testing experimental phase, the corresponding data set consists of coordinates of X, Y, Z, Roll, Pitch and Yaw under different light intensity as well as different distances between transmitter and Head Tracker (receiver). Accuracy and sensitivity of predicted 6-DoF data using different machine learning methods have been described. The corresponding data may be low in features but it is high in observation values. Generalized results have been provided in Table 5.

| Model | Distance | Light Intensity |
|---|---|---|
|  | Accuracy | Accuracy |
| R Forest | 99.05 | 98.03 |
| SVM | 93.21 | 92.22 |
| LM | 71.49 | 70.95 |
| NN | 51.33 | 52.11 |

Table-V: Comparison of performance of different methods of machine learning for prediction of 6-DOF data under different light intensity and distances between transmitter and receiver

### 4.1 Training-Testing Experiment

In training-testing phase, the distribution of data are set to [50% and 50%], [60% and 40%], [70% and 30%] and [80% and 20%] respectively for all four machine learning methods (the corresponding data set used in the experiment can be found in the package). Table 6 and Table 7 provide describe comparison of performances of all the machine learning methods in which 6-DoF data is predicted under different light intensity and distance range in terms of accuracy.

| Models | Training testing partition | | | |
|---|---|---|---|---|
|  | 50-50% | 60-40% | 70-30% | 80-20% |
| R Forest | 90.56 | 98.78 | 98.03 | 99.25 |
| Linear Model | 71.77 | 72.88 | 70.95 | 71.56 |
| SVM | 92.34 | 90.32 | 92.22 | 94.40 |
| NN | 51.12 | 54.43 | 55.11 | 56.31 |

Table-VI: Training and testing partition for the prediction of 6-DoF data under different light intensities

| Distance | Training Testing Partition | | | |
|---|---|---|---|---|
| Models | 50-50% | 60-40% | 70-30% | 80-20% |
| Random Forest | 98.74 | 99.05 | 99.05 | 97.57 |
| Linear Model | 72.33 | 70.46 | 71.49 | 72.08 |
| SVM | 87.14 | 89.30 | 93.21 | 93.20 |
| NN | 50.15 | 51.13 | 55.83 | 56.64 |

Table VII: Training and testing partition for the prediction of 6-DoF data under different distances between receiver and transmitter

The performance comparison of different machine learning concludes that Random forest method has better accuracy over other machine learning methods used for prediction. Table 5 depicts the comparative performance of all the four methods of machine learning in the prediction of 6-DoF data under the different light intensity and distances in terms of accuracy.

5. Conclusion

In this letter, four different machine learning methods (random forest, linear model, support vector machines and neural network) over experimentations with the six movements of head in an optical head tracking have been explored on cockpit display simulator. Further, the 6-DoF data is predicted using machine learning methods under different light intensity and distances between transmitter and receiver. The experiments were conducted under different atmospheric conditions like different light intensities and different distance between transmitter and receiver. This is done to predict the working of an optical tracker under various atmospheric conditions. With the prediction percentage achieved by the various machine learning methods it may help the pilot to choose the appropriate tracker for the tracking of head of the pilot.

6. Future Scope

Similar study can be carried out for different tracking technologies, so that the machine learning predictions could help in choosing the best tracking technology. The result may be single tracking technology or the hybridization of different technologies.

**Acknowledgment**

We thank the CSIR-CSIO Laboratory for the use of their equipments and the support to conduct experiments.